\documentclass[twocolumn]{aastex631}

\usepackage{float}

\newcommand{\leda}{LEDA\;1313424}
\newcommand{\kms}{km s$^{-1}$}

\usepackage{graphicx}
\graphicspath{{./}{figures/}}

\begin{document}

\title{The Bullseye: HST, Keck/KCWI, and Dragonfly Characterization of a Giant Nine-Ringed Galaxy}

\author[0000-0002-7075-9931]{Imad Pasha}
\affiliation{Department of Astronomy, Yale University, 219 Prospect Street, New Haven, CT 06511}

\author[0000-0002-8282-9888]{Pieter G. van Dokkum}
\affiliation{Department of Astronomy, Yale University, 219 Prospect Street, New Haven, CT 06511}

\author[0000-0002-7490-5991]{Qing Liu}
\affiliation{Department of Astronomy,University of Toronto}

\author[0000-0003-4381-5245]{William P. Bowman}
\affiliation{Department of Astronomy, Yale University, 219 Prospect Street, New Haven, CT 06511}

\author[0000-0003-0327-3322]{Steven R. Janssens}
\affil{Centre for Astrophysics and Supercomputing, Swinburne University, Hawthorn VIC 3122, Australia}

\author[0000-0002-7743-2501]{Michael A. Keim}
\affil{Department of Astronomy, Yale University, New Haven, CT 06520, USA}

\author[0000-0002-6558-9894]{Chloe Neufeld}
\affiliation{Department of Astronomy, Yale University, 219 Prospect Street, New Haven, CT 06511}

\author[0000-0002-4542-921X]{Roberto Abraham}
\affiliation{Department of Astronomy,University of Toronto}

\begin{abstract}
We report the discovery and multiwavelength followup of LEDA 1313424 (``Bullseye''), a collisional ring galaxy (CRG) with nine readily-identified rings --- the most so far reported for a CRG. These data shed new light on the rapid, multi-ring phase of CRG evolution. Using Hubble Space Telescope (HST) imaging, we identify and measure nine ring structures, several of which are ``piled up'' near the center of the galaxy, while others extend to tens of kpc scales. We also identify faint patches of emission at large radii ($\sim$70 kpc) in the HST imaging, and confirm the association of this emission with the galaxy via spectroscopy. Deep ground based imaging using the Dragonfly Telephoto Array finds evidence that this patch of emission is part of an older, fading ring from the collision. We find that the locations of the detected rings are an excellent match to predictions from analytic theory, if the galaxy was a 10-ring system whose outermost ring has faded away. We identify the likely impacting galaxy via Keck/KCWI spectroscopy, finding evidence for gas extending between it and the Bullseye. The overall size of this galaxy rivals that of known Giant Low Surface Brightness Galaxies (GLSBs) such as Malin I, lending credence to the hypothesis that CRGs can evolve into GLSBs as their rings expand and fade. Analysis of the HI content in this galaxy from ALFALFA finds significantly elevated neutral hydrogen with respect to the galaxy's stellar mass, another feature in alignment with GLSB systems. 
\end{abstract}

\keywords{Low surface brightness galaxies (940), Giant galaxies (652), Ring Galaxies (1400)}

\section{Introduction} \label{sec:intro}

\begin{figure*}
    \centering
\includegraphics[width=\linewidth]{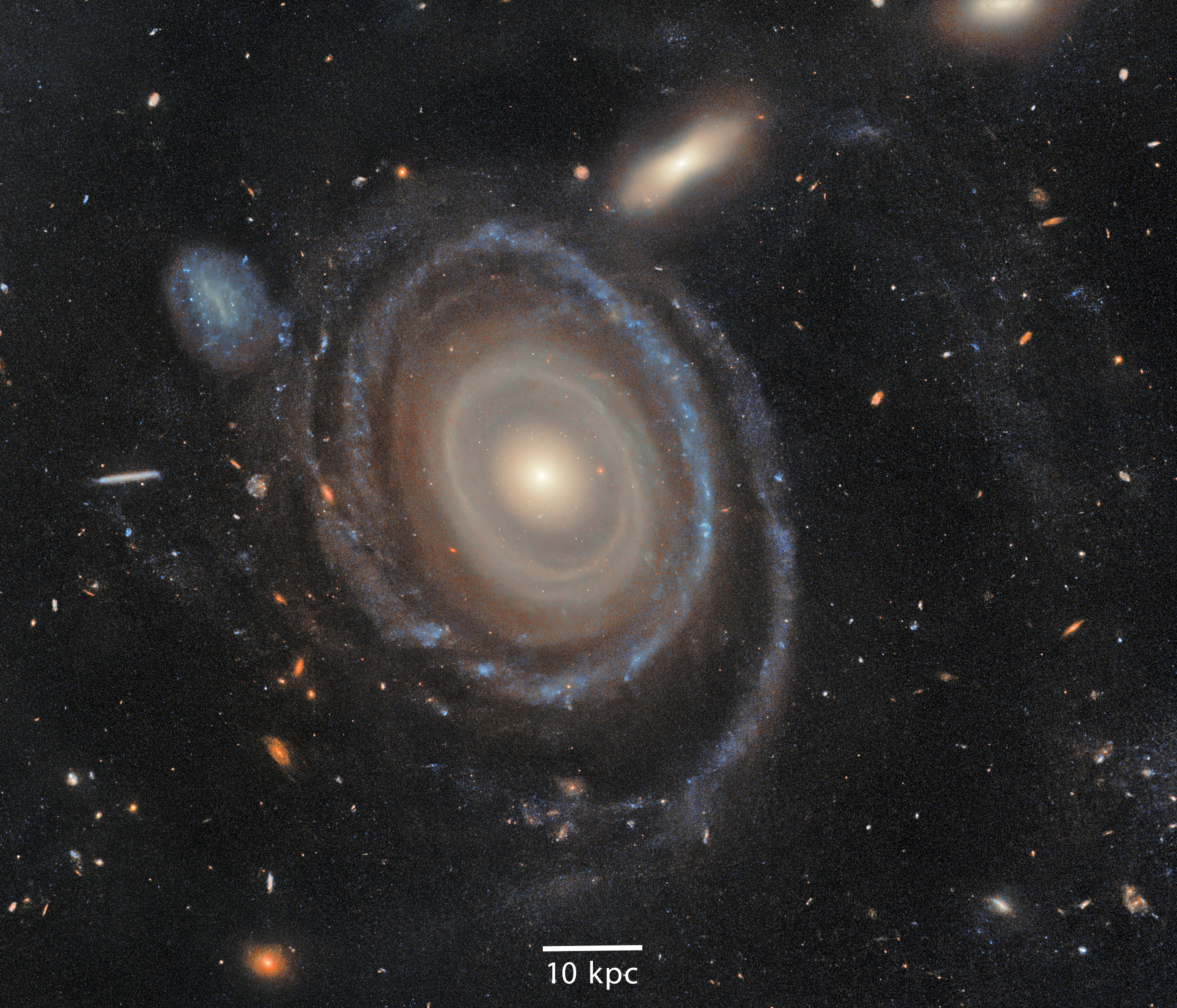}
    \caption{HST image of \leda{} (the ``Bullseye''), processed by \textit{ESA}, with additional cropping and contrast adjustments made to highlight some of the low surface brightness structures. We do not use this image for any measurements, but it provides a striking view of the varied rings throughout this system, from red inner rings (where gas may have been evacuated), to blue, highly star-forming rings, to evidence west of the galaxy for a fading, expanded ring at large radius (which have since been spectroscopically confirmed). That faint material extends up to $\sim70$ kpc from the center of the galaxy. The image also captures a clumpy, blue galaxy which appears to have strong star-bursting behavior and which has a line-of-sight velocity a few hundred km s$^{-1}$ offset from the Bullseye. While not shown in this work, KCWI spectroscopy of the Bullseye core finds elevated [NII]/$H\alpha$ ratios suggestive of AGN triggering.}
    \label{fig:esa}
\end{figure*}

Collisional Ring Galaxies (CRGs) have been long known, due to the nearby Cartwheel Galaxy's identification as a candidate CRG by Zwicky in the early 1940s \citep{Zwicky:1941}. These systems, first theoretically described in a series of papers \citep[e.g.,][]{Lynds:1976,Theys:1976,Theys:1977} based on the earliest set of known examples (e.g., II Hz 4), form when an impactor galaxy ``drops through'' the center of a (typically more massive) disk galaxy. When this occurs as a near head-on collision through the center, the density perturbation induced can produce rings which expand outward, sweeping up gas along the way, often evacuating the host galaxy's central regions of gas while triggering starburst-like conditions within the rings \citep[e.g.][]{Struck-Marcell:1987,Renaud:2018}. 

As wider field surveys produced larger samples of imaging data, e.g., the Sloan Digital Sky Survey, further work expanded on both the theoretical and observational fronts \citep[e.g.,][]{Hernquist:1993,Marston:1995,Appleton:1997}, and then again more recently as surveys both increased in coverage and depth as well as were supplemented by citizen scientists to visually identify samples of rings \citep[e.g., Galaxy Zoo;][]{Lintott:2008}, which led to the re-examination and characterization of these systems \citep{Madore:2009,Struck:2010}.

CRGs are valuable laboratories for understanding galactic structure and evolution, as the perturbative nature of the collision and resultant expanding rings provide sensitive probes of both the shape and strength of the gravitational potential of the impacted galaxy, as well as the processes driving star formation in propagating density waves. They are also important for our understanding of galaxy mergers and interactions, as they represent a rare, highly symmetric case where an analytic theory can make tractable and testable predictions about a system's time evolution. Furthermore, and of particular importance for the work that will be presented in this paper, there is some evidence that after the brief ring phase following a collision, the expanded and faded rings may result in systems resembling giant low surface brightness galaxies (GLSBs), observationally-rare objects that are challenging to explain in standard $\Lambda$CDM physics \citep{Mapelli:2008}.

Unfortunately, the ring-dominated phase of post-collision evolution for CRGs is exceedingly short by cosmic standards: simulations broadly find that for different impactor scenarios (e.g., mass ratios, velocity differences) rings tend to dissolve after a few hundred Myr \citep[e.g.,][]{Struck:2010}. To date, no CRGs have been identified containing more than three rings, and the majority of two or three ring candidates that have been identified have been challenging to study; often their distance means they are small in angular size and under-resolved in ground based imaging surveys for the type of detailed analysis required to vet analytic theory \citep[e.g.,][]{Madore:2009}.

\leda, (which we have nicknamed the ``Bullseye'') was discovered serendipitously in Legacy Survey DR9 imaging \citep{Dey:2019}. The galaxy has a redshift of $z=0.039414$, corresponding to a luminosity distance of 173.9 Mpc for $H_0=70$\,km\,s$^{-1}$\,Mpc$^{-1}$. Upon visual inspection, we identified the galaxy as a multi-ring galaxy, with two outer highly star forming rings enclosing much redder inner ring structures. Lower signal-to-noise ring-like arcs were identified between the more dominant rings, and faint, blue emission was detected in the far outskirts of the system, at galactocentric radii of $\sim70$ kpc. Followup multiband HST/ACS imaging of the galaxy was obtained in F475W and F814W as part of GO 17508 in Cycle 31 (PI: Pasha). A color image created from the HST data by the European Space Agency (\textit{ESA}) is presented in Figure \ref{fig:esa}, highlighting the remarkable morphology of this galaxy. Beyond its multi-ring structure, it also contains a multitude of point sources in the inner galaxy (which may be star clusters) and an elliptical-like, $r^{1/4}$ core, not dissimilar to Hoag's Object \citep{Hoag:1950} or PGC 1000714 \citep{MutluPakdil:2017}. Surface brightness profiles show no sign of a bar, though no observations have as yet been made in the restframe near-infrared. Spectroscopy reveals AGN-like line ratios in its center, likely triggered by the collision.

Beyond the galaxy itself, two systems at similar redshifts were identified, with one being the most probable impactor. In Figure \ref{fig:hst-1}, we show the authors' color image combination created using the science-calibrated imaging and placed on to the world coordinate system, with a followup KCWI pointing shown (see \S\,2.2.1). 

In this Letter, we present a multi-instrument study of the system, finding evidence that the collision created  10 rings, an unprecedented nine of which are detected (and one of which is assumed to be faded away at extremely large radii; see Section \ref{sec:many-rings}). While the ring characterization is carried out primarily using HST imaging for the innermost rings, we also identify and characterize an outer ring at large galactocentric radius via low surface brightness-optimized imaging from the Dragonfly Telephoto Array, confirmed spectroscopically to be associated with the Bullseye via Keck/KCWI. Additionally, we present Keck/KCWI spectroscopy which strongly suggests the blue, clumpy system in Figure \ref{fig:esa} is the impacting galaxy. 

Finally, we discuss the late stage evolution of this system, in light of its spatial extent out to radii approaching those of giant low surface brightness galaxies (GLSBs; e.g., Malin I), and theoretical predictions that such a transition from CRGs to GLSBs may be possible \citep[e.g.,][]{Mapelli:2008}. We find that not only does this object have optical properties consistent with a transition from CRG to GLSB, its neutral hydrogen properties also match the elevated values found in such systems.

\section{Ring Identification and Characterization}\label{sec:many-rings}

This theory of ring formation and evolution has been developed since \cite{Appleton:1987}; here we use the practical model proposed in \cite{Struck:2010} to asses the rings in \leda{}. 

Practically, in the perturbative limit and for flat rotation curves, the analytic theory makes a remarkably simple (and testable) prediction for the structure of galactic rings formed post-collisions, namely, that 

\begin{equation}\label{eqn:radii}
\frac{r_{i}}{r_{i+1}} = \frac{2i+1}{2i-1},
\end{equation}
where $i$ is the ring number and $r$ is the radius of the $i$th ring.

Equation 1 predicts that the ratio of sizes between the first- and second-created rings should be 3.0, with successive ratios for rings 2-10 being 1.67, 1.40, 1.29, 1.22, 1.18, 1.15, 1.13, and 1.11.

To date, very few systems have been found to test this theory; by far the most common example of ring galaxies include only a single ring. A small number ($\sim$dozens) of examples of double (or tentatively triple) ring systems were used to test the analytic ring theory in \cite{Struck:2010}. The ratios reported in that work were all between 2$-$3.5, which place those systems firmly in the range of tracing rings 1-2 or 2-3.  To date, no CRG with more than three well-defined rings has been reported. 

As a note, in this Letter we work primarily using ellipses, quoting the semi-major axis when we refer to a radius. The inclination of the system (assuming the rings are $\sim$circular intrinsically), is measured to be
\begin{equation}
    i = (41.9 \pm 1.5)^{\circ}, 
\end{equation}
taking the average and standard deviation of four measurements (three from HST and one from Dragonfly); this implies an axis ratio correction of $\sim$0.75. 

\subsection{Inner Rings}

\begin{figure}
    \centering
    \includegraphics[width=\linewidth]{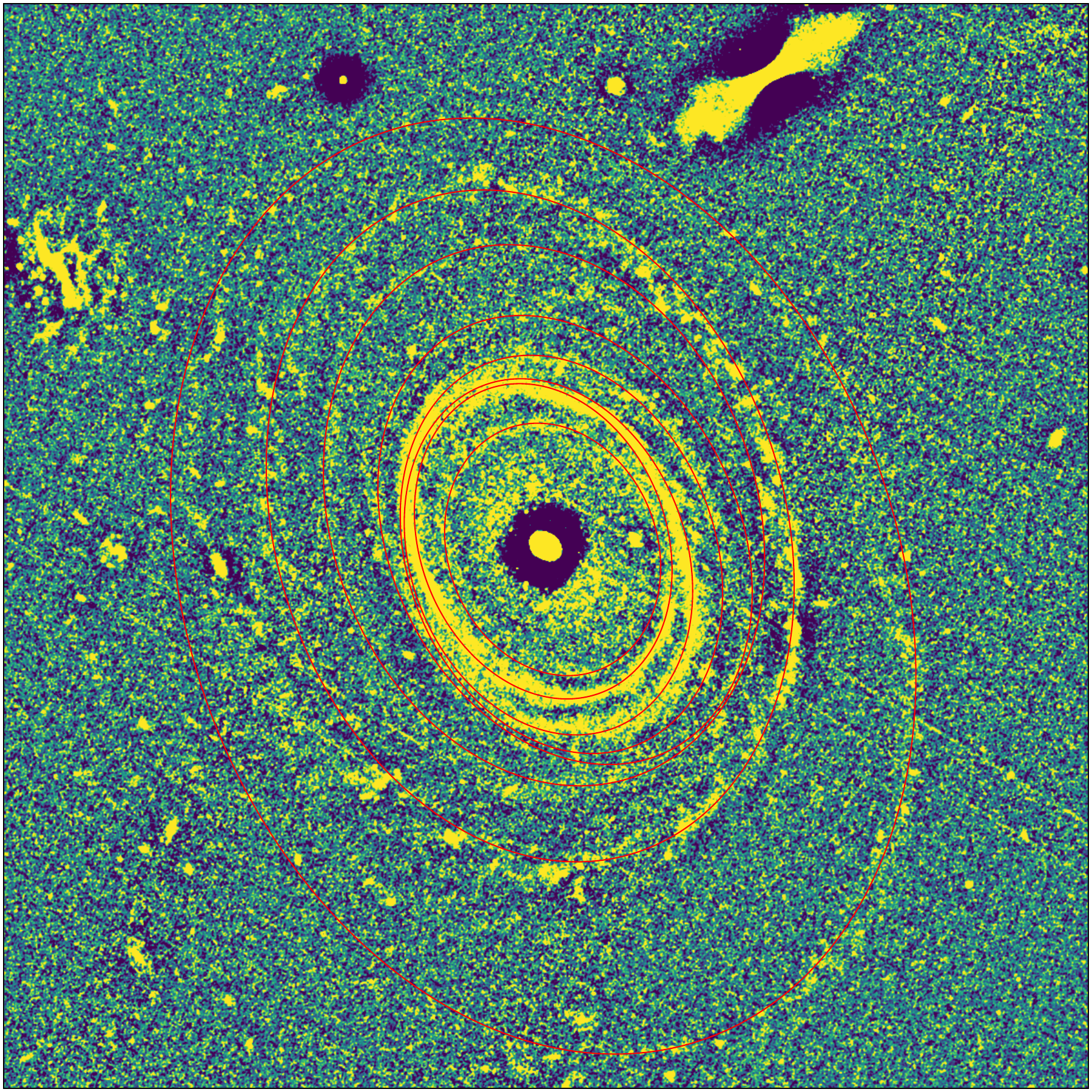}
    \caption{Unsharp-mask view of the galaxy, with red ellipses overplotted over structures initially identified as rings. Restricted-radius radial profiles were then used in the vicinity of each structure to measure their centroids.}
    \label{fig:ring-initial}
\end{figure}

\begin{figure*}
    \centering    \includegraphics[width=0.495\linewidth]{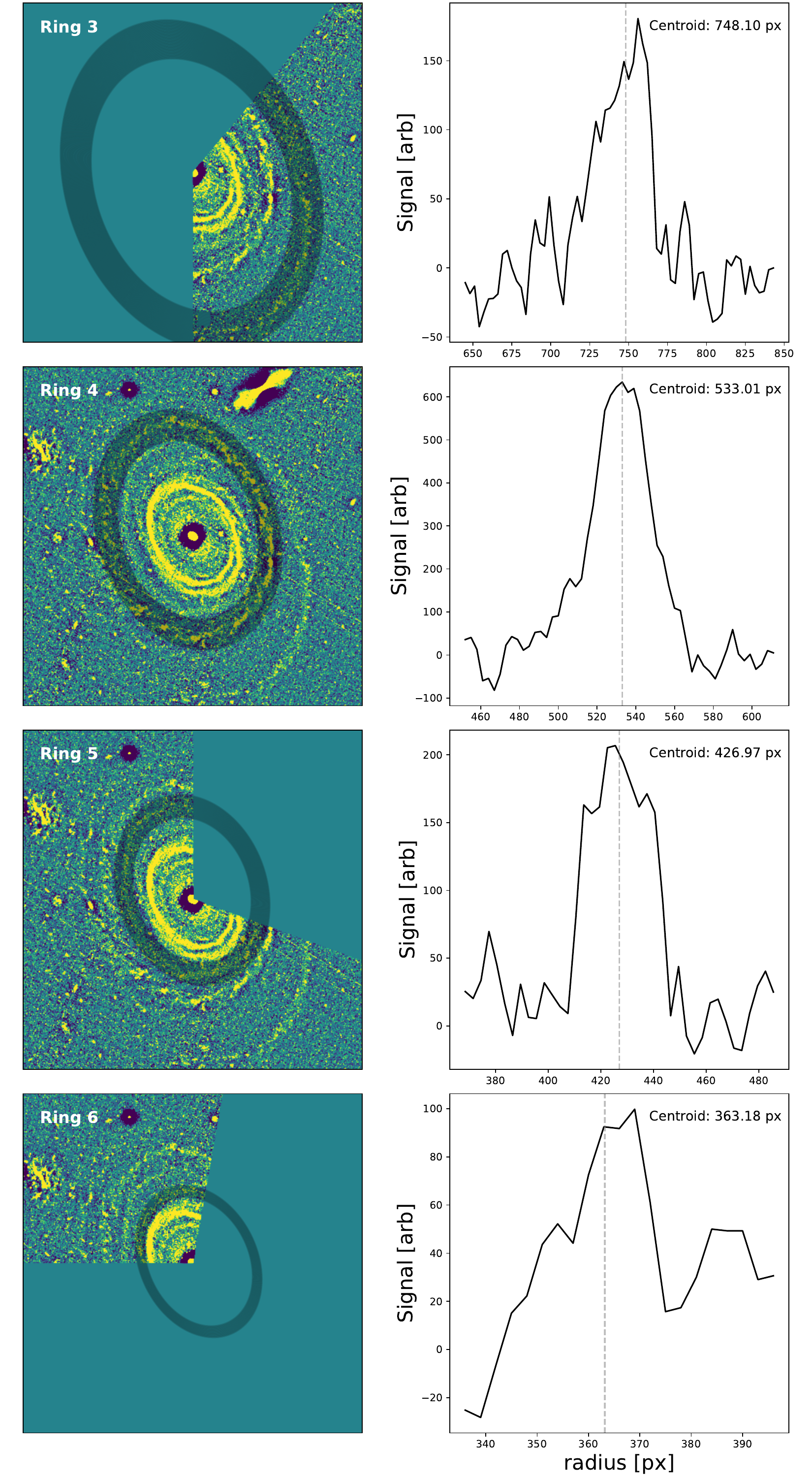}
    \includegraphics[width=0.495\linewidth]{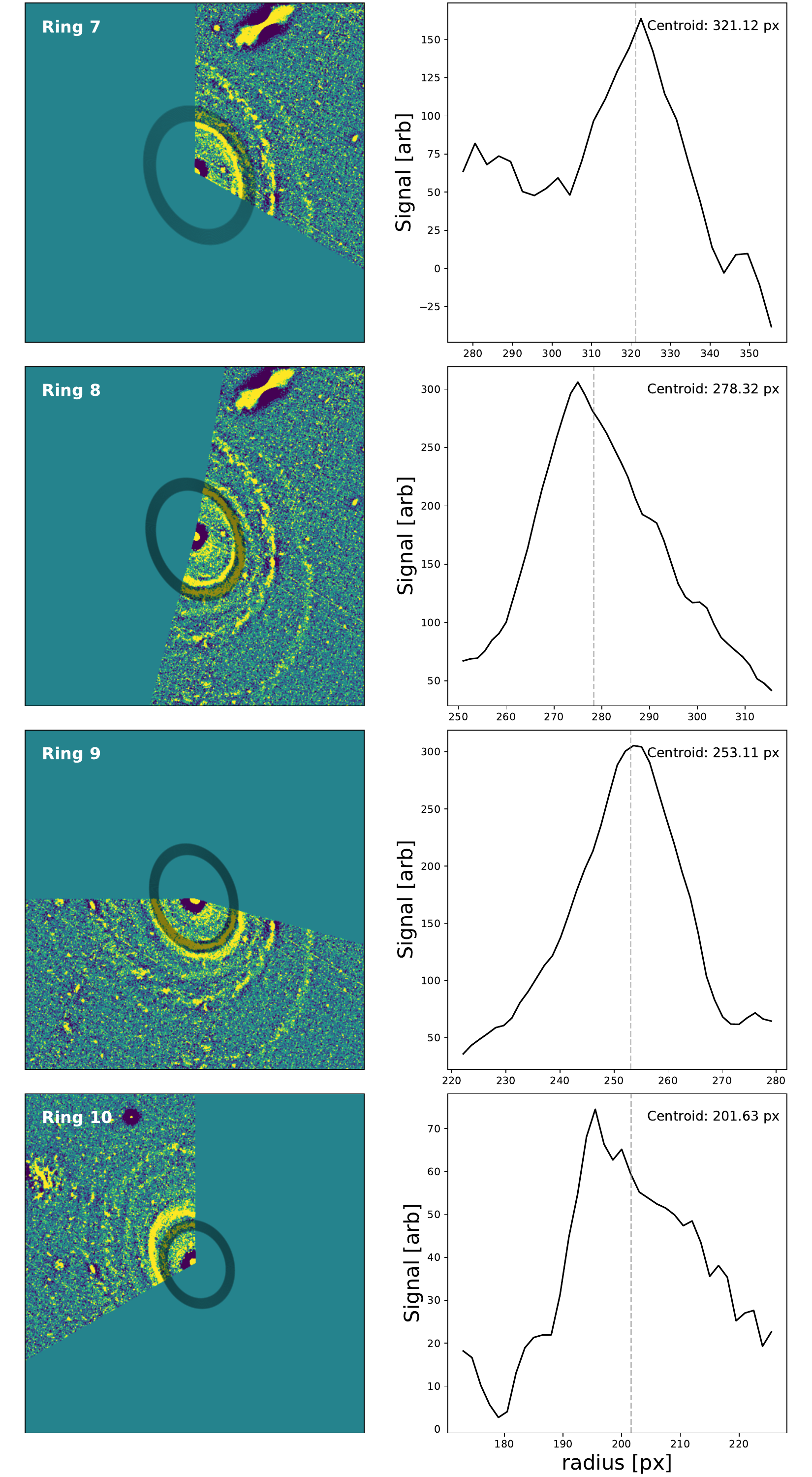}
    \caption{Overview of the Ring characterization procedure for Rings 3 through 10. In the left-columns, we show the unsharp mask image of the inner galaxy, masked to the span in $\theta$ over which we determined the profile if relevant. Overplotted are the annuli  used for construction of the profile. In the right columns are the extracted profiles, with centroids denoted as calculated from within the FWHM of each peak. }
    \label{fig:cen-1}
\end{figure*}

\begin{figure*}
    \centering
    \includegraphics[width=\linewidth]{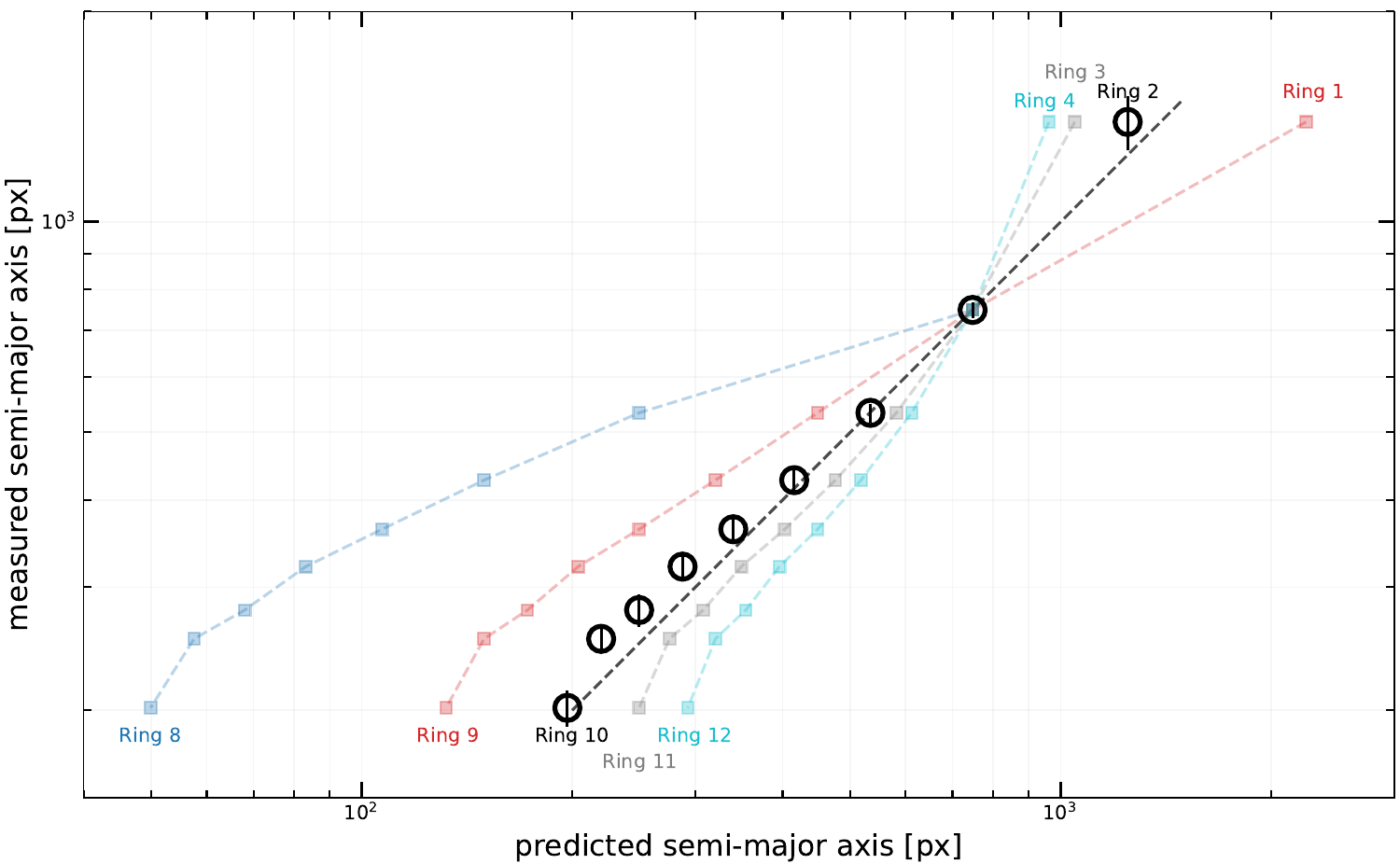}
    \caption{Analytic ring theory makes predictions for the \textit{ratio} of successive ring radii; thus, if one assumes a particular ring number for a particular measured ring, all other ring positions can be predicted. Here, we show the measured ring radii as a function of such a prediction, in which the outermost narrow star-forming ring is assumed to be Ring 3 (black), or several other possible ring numbers (1, 2, 4,  and 5) in colors. From this comparison, it is clear that the interpretation of the outermost narrow star-forming ring as Ring 3 is the most consistent with the data; the black dotted line shows the 1:1 relation. It is further clear that the predictions in this case are remarkably close to the measured values across all detected rings. Note that we detect a structure (Ring 2) beyond the outermost narrow star-forming ring, which precludes it being Ring 1, via Dragonfly Telephoto Array imaging (discussed in \S \ref{subsec:dragonfly}). In all cases, the predicted ratios are based on an assumption of a flat rotation curve; in this case, inner rings (beyond 3) are expected to overlap, thus, the offset from the 1:1 relation for some inner rings could be a result of measuring the outer or inner edges of broad, overlapping rings.}
    \label{fig:ring-compare}
\end{figure*}

\begin{figure*}
    \centering
    \includegraphics[width=\linewidth]{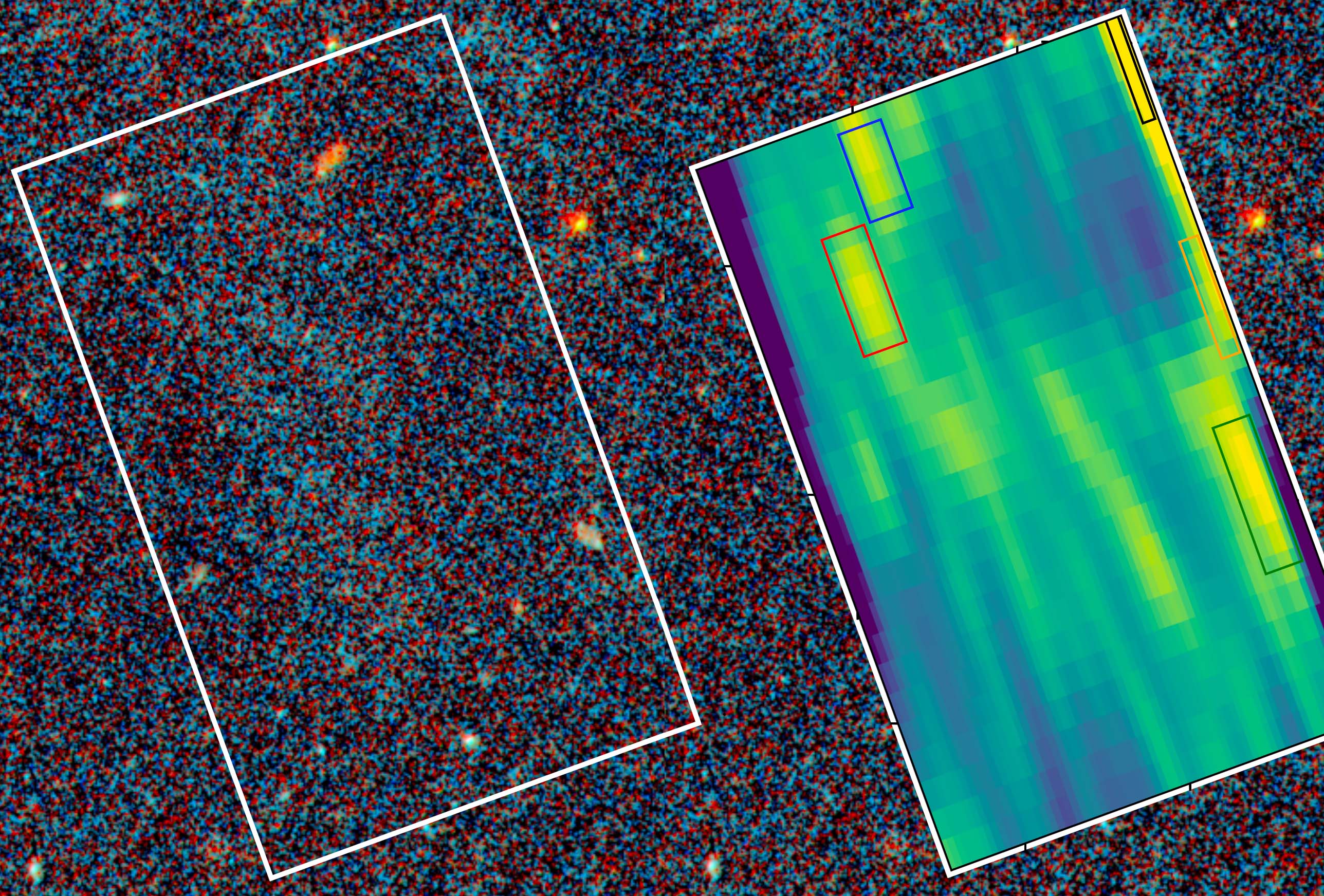}
    \caption{Zoom in of the region targeted by KCWI, with the large slicer (20.4$''$ $\times$ 33$''$) targeted field of view shown at left. On the right, we show a moment 0 map of the KCWI data cube constructed for a narrow velocity range around the $\sim$11816 km s$^{-1}$ recession velocity of \leda, smoothed with a gaussian kernel with $\sigma=$1.8. In white boxes, we show several regions with identified emission line detections with velocities consistent with \leda; these boxes are replicated on the left to show from where the emission is originating. One-dimensional spectra extracted from these regions can be found in Figure \ref{fig:one-d-spec}. } 
    \label{fig:kcwi-overlay}
\end{figure*}

We use the high resolution HST imaging to characterize the inner rings that ``pile up'' within the two star-forming rings. In Figure \ref{fig:ring-initial}, we show an unsharp-mask version of the HST image (combining both bands) for the inner $\sim$arcmin of the galaxy, created using the \texttt{skimage} library. Beginning with the outer (star forming) ring, we visually identified ring-like structures, assigning an ellipse to every significant arc of emission (shown in red ellipses in Figure \ref{fig:ring-initial}). 

In total, we identify eight rings or ring portions in the data, including the outer star forming ring, with progressively smaller ratios in radius. All identified ring structures share similar position angles ($\sim20$ degrees) and axis ratios ($\sim0.75$), though the center varies between ellipses. Because some of the rings have only partial arcs and low signal to noise, formal fitting procedures and uncertainty estimations struggle to constrain the ellipses needed to fit each ring. We thus adopt the following procedure: 
\begin{enumerate}
    \item First, ellipses are matched to each ring structure by eye; here position angle and axis ratio were allowed to vary along with position, but in practice the axis ratio was nearly 0.75 in all cases, while $\theta$ varied from 19-25 degrees. 
    \item Next, elliptical annuli which match the center, position angle, and axis ratio are placed in a window extending radially from the initial estimates.
    \item For each ring, an arc in position angle is used to characterize the ring, chosen to maximize the region of the ring that is most-separated from adjacent rings.
    \item A radial profile is computed from the annular apertures, within the radial window. On occasion a particularly bright blob of emission is masked so as to not bias the overall ring determination.
    \item Finally, the peaks in the resulting 1D profiles are characterized by a centroid computed within the full-width half-maximum (FWHM) of the peak, with that FWHM taken as the nominal uncertainty on the derived semi-major axis.
\end{enumerate}

In Figure \ref{fig:cen-1} we show the outcome of this process for all identified rings. In the left columns, we show the unsharp mask image with the subsection in $\theta$ used to determine the profile, if relevant, with the annuli range used to cover the ring and its immediate background shown in black. In the right columns we show the profiles extracted from these annuli, with the centroids (calculated within the FWHM of each peak) denoted. We adopt these centroids as our nominal semi-major axis values for each ring (note that the abscissa of each profile is the semi-major axis of the annulus within which the flux at that position was computed). Ring 6 has a somewhat double-peaked structure; we use the inner of these structures based on visual inspection.

We identify the number of each ring (in order of creation time) by utilizing the ratios predicted for the radii of rings by analytic theory. The two well-defined, blue star forming rings have a radial ratio of nearly exactly $1.4$, the predicted ratio between the third and fourth created ring. This then guides the numbering scheme we utilize throughout the paper; rings progressively inward are given higher ring numbers. As is shown in Figure \ref{fig:ring-compare}, comparisons across all detected structures further indicate that the outer star-forming ring is indeed Ring 3.

Based on this determination, we extrapolate via Equation \ref{eqn:radii} the semi-major axes expected for rings 4-10 (working inward from the outer star forming ring), and for rings 2 and 1 (working outward) and compare the predicted ring semi-major axes from the analytic theory (given the measured semi-major axis for ``Ring 3'') to the semi-major axes derived from the centroids of the detected peaks.

We show the results of this comparison in Figure \ref{fig:ring-compare}. The rings in \leda{} show remarkable agreement with the analytic theory, in the first test of the continued adherence to the predicted ring ratios past 2 or 3 rings. Note that we add a datum for Ring 2 to this figure, which is derived in \S 2.2.2. We find no direct evidence for Ring 1; at 3 times larger distance, an expected surface brightness fainter than Ring 2, and in a field of significant cirrus contamination, we do not expect to detect it. 

For reference, we also show the predicted vs. observed relations under other assumptions for which ring number corresponds to the outer star forming ring; beyond the near-perfect agreement in ratio between the two star forming rings (1.4) for a ring 3 determination, \textit{all} predicted radii are significantly more offset from a 1:1 match for any other determination of that ring, with it being impossible to be Ring 1 (see \S 2.2.1 and \S 2.2.2) and far less likely to actually be ring 2, 5, or 6. The most viable alternate interpretation is that the outer star forming ring is ring 4, rather than 3, which implies the inner-most ring we detect would be Ring 11. However, as we will show in \S 2.2.1 and \S 2.2.2, we detect and can roughly characterize a ring beyond the outer star-forming ring, which provides a strong lever arm; as shown in Figure \ref{fig:ring-compare}, the addition of this datum strongly constrains the scenario at hand to being detections of Ring 2 - Ring 10 in this system.

While arbitrarily many further (smaller, newer) rings can be predicted from Eqn. 1, we see no obvious signs of further ring structures within Ring 10 (though we note that the steep light gradient and presence of the bulge-like center would make it considerably challenging to identify further rings if present). We thus infer that the 9 observed rings are likely the result of a collision that originally produced at least 10 rings, with the remnants of Ring 1 expected to be $\sim3$ times further out than the detected Ring 2 (and likely wholly undetectable).

The predictions for the ring ratios used here assume a flat rotation curve (FRC); however, the FRC case makes the additional prediction that beyond ring $N\sim 3$ (i.e., inner rings), the rings should broaden and overlap, generally to the point of indistinguishability \citep[e.g.,][]{Appleton:1996}. There are several ways to comport this prediction with the observed rings. 

First, the theoretical ratio prediction in the FRC case can apply to the inner or outer edge of rings as well as their centers. In the case of broad, overlapping rings, it is more likely to detect the edges of the ring structures than their centers; thus, the inner measurements here may represent the inner or outer edges of inner rings (or a mix). This is an interesting possibility, given the (slight) offset of rings 6,7,8 and 9 above the relation predicted from ring 3 (which is definitively coming from a ring center).

Alternatively, the center of the galaxy could have a rising rotation curve, in which case the rings can be narrow, as the radial derivative of the epicyclic frequency is small. Future work (particularly, the measurement of a rotation curve for the system) can help distinguish these scenarios. 

Similarly, there are structures present in the disk which are not purely ring-like; in particular, certain windings, incomplete arcs, and splits in structure could also resemble spiral structure. Some of these structures could reflect the progenitor's spiral structure, not yet completely re-arranged. There are also scenarios for the collision --- particularly, a prograde component to the impactor's orbit --- which produce waves which resemble narrowly-wound spirals that can resemble rings \citep{Struck:2011}. However, we find this interpretation is less likely to describe the Bullseye given the longer timescales involved; the spirals induced require $\sim$several rotation periods ($\sim$hundreds of Myr) to develop, while we find a significantly shorter time since collision for \leda{} (see \S 3). 

\subsection{Outer Rings}
Here we investigate the emission seen at large radii, which the analytic theory predicts to be the expanded, faded second ring. We first assess whether there is a robust confirmation that at least some of the emission seen is associated with \leda{} and is not, e.g., Galactic cirrus, as the Bullseye is in a region of fairly significant cirrus contamination. 

\subsubsection{Keck/KCWI Spectroscopy}

Using 8 minutes of on-source Keck/KCWI spectra obtained of October 20, 2023 using the large slicer and RH2 grating on the patch of emission identified in Figure \ref{fig:hst-1} and \ref{fig:kcwi-overlay}, we search for emission line detections from residual ionization of the gas at the redshift of \leda. Such detections would be expected if the emission is indeed faded rings, as the outer rings will have been strongly star-forming in the recent past and may still have residual ongoing star formation.

\begin{figure}
    \centering
    \includegraphics[width=\linewidth]{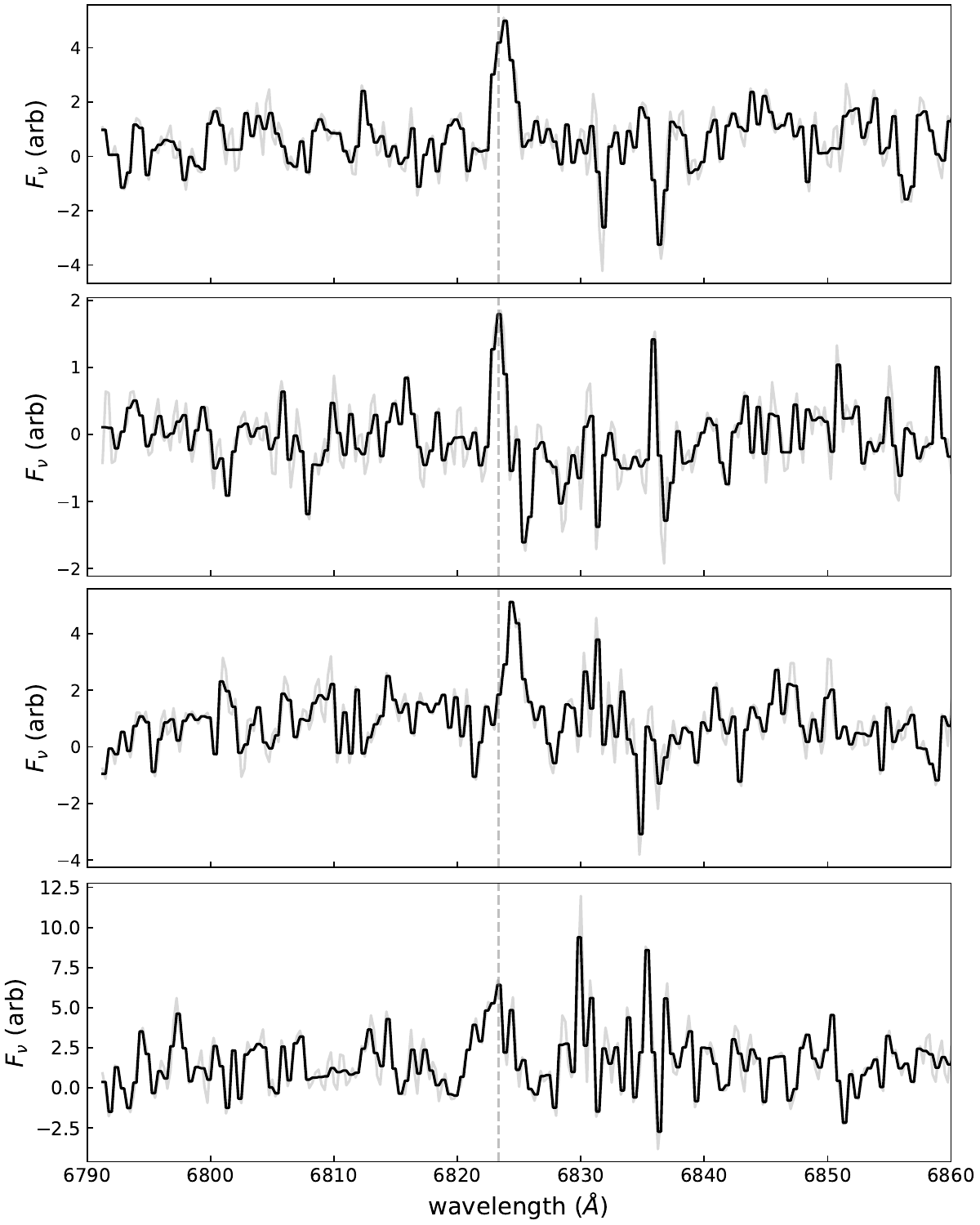}
    \caption{Mean-collapsed one-dimensional spectra extracted from the regions identified in Figure \ref{fig:kcwi-overlay}, in the observed (heliocentric, vacuum) frame. Raw extracted spectra are shown in grey; a binned spectrum (2 or three-element average) is shown in black, with bad pixels masked, and a vertical dashed line indicates the expected $H\alpha$ location for gas with an identical recession velocity to \leda. We find several clear detections of $H\alpha$ emission within the KCWI footprint over the patch of emission in the outskirts of the Bullseye, which span a small velocity range around the galaxy's recession velocity.}
    \label{fig:one-d-spec}
\end{figure}

\begin{figure*}[htp]
    \centering
    \includegraphics[width=\linewidth]{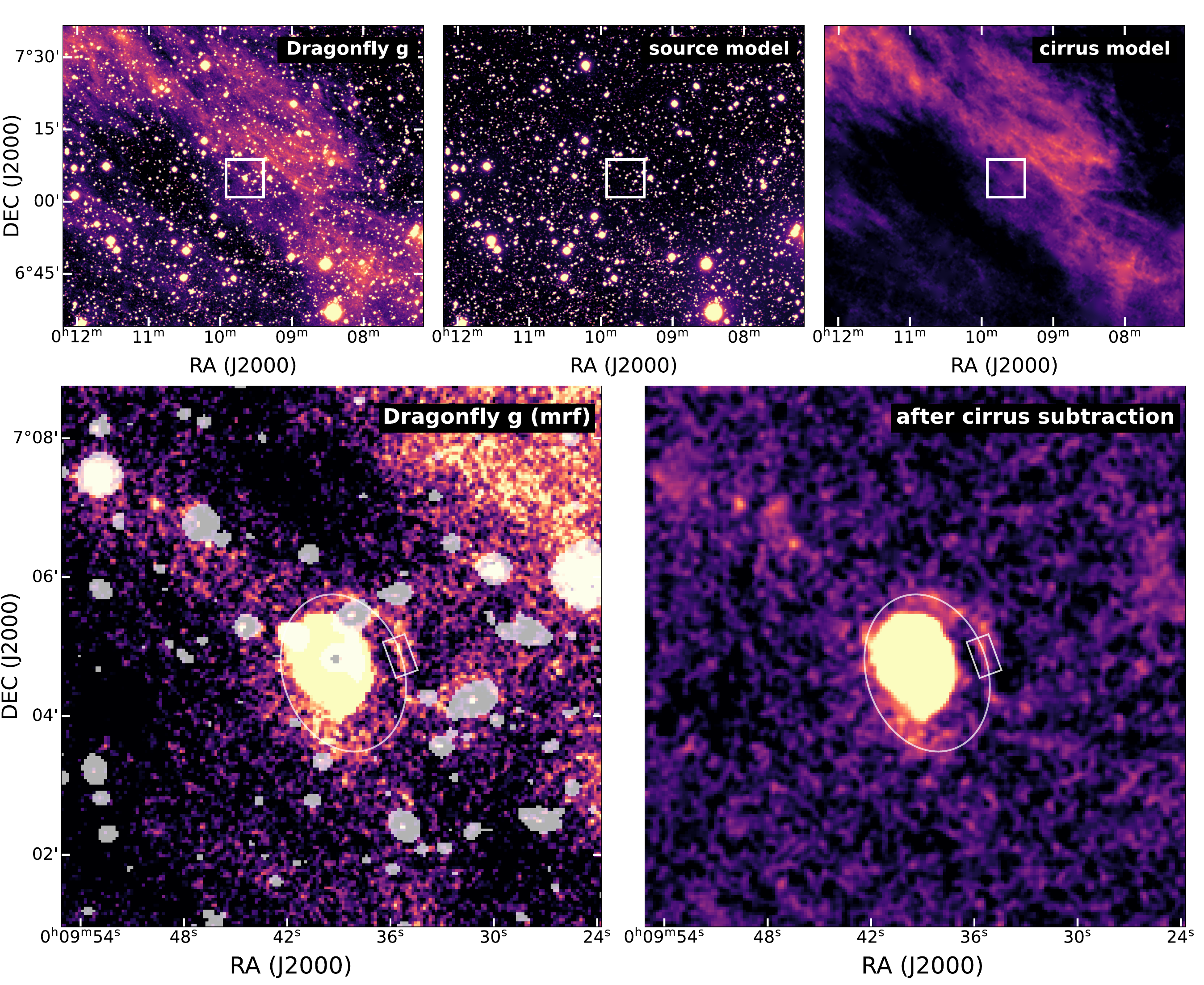}
    \caption{Summary of the cirrus modeling and subtraction procedure. In the top row, we show wide field view of the region, with the original Dragonfly $g$-band image (left), source flux model constructed by \texttt{mrf} (center), and cirrus model created using the MRF-subtracted image (right). A white box in each panel identifies the location of \leda{} in the field. In the bottom row, we show zoomed in panels of the apertures shown in the top row, of the \texttt{mrf}-filtered $g$-band image (left) and residual after cirrus modeling, subtraction, and smoothing (right), both with an overlay of the KCWI pointing in white within which the H$\alpha$ detections were made at the redshift of \leda. Residuals from \texttt{mrf} subtraction were masked; these mask locations are shown in the left-hand image as semi-transparent overlays. The \texttt{maskfill} package was used to fill in these source masks on the right. We find clear evidence for Ring 2, which shares a similar position angle and axis ratio to the inner rings. We mark a nominal matched ellipse to this ring structure in white. }
    \label{fig:subtracted}
\end{figure*}

The recession velocity of \leda{} is 11816 $\pm$ 248 km s$^{-1}$ via the ALFALFA spectrum. We identify emission line detections in the KCWI cube in multiple slices. One detection (in a single slice) is at a clear offset from the Bullseye, roughly 1400 km s$^{-1}$ blueward of the galaxy. The rest of the detections, spanning $\sim$7 slices, are all found to be at the redshift of \leda: these detections are all within 40 km s$^{-1}$ of the nominal galaxy systemic velocity, and have a narrow spread (of order 40 km s$^{-1}$), firmly tying them to the galaxy and to each other.\footnote{We measure the recession velocity of \leda{} to be 11,731 km s$^{-1}$ with KCWI in \S 3; these KCWI data were obtained at a different time and do not share calibrations, so we defer to the ALFALFA velocity; still, either would produce consistency with this emission's association with the Bullseye. } 

In Figure \ref{fig:kcwi-overlay}, we show a zoom in on the region targeted by our KCWI observations, with a moment 0 map computed for a narrow velocity range around the systemic velocity of the Bullseye overlaid on the footprint. We identify several regions within the cube footprint with $H\alpha$ detections, for which we extract 1-D spectra (shown in Figure \ref{fig:one-d-spec}). As expected, these detections come from blue, clumpy regions within the gas; one detection (more marginal) at the very edge of the cube appears to be the edge of a clump of blue emission that extends away from the footprint. 

The robust line detections in the 1-D spectra confirm the peaks in the moment map are not spurious, and confirm unambiguously that there are in fact clumps of gas at large ($r>70$ kpc) galactocentric radii in this system. The material targeted by these KCWI observations should be part of Ring 2, according to analytic theory, which given our Ring 3 determination is expected to have a semi-major axis of 62.1$''$. Assuming the same position angle and axis ratio as ring three, this generally places the prediction just within the inner edge of the seen emission. It is challenging to constrain just how large the discrepancy is with the HST data alone as the identified emission cannot be robustly fit, and thus small changes in center, position angle, and axis ratio of a presumed model ring could change the offset considerably; we return to this question in the \S 2.2.2, where Dragonfly data provide better constraints on the observed structure.

\subsubsection{Dragonfly Imaging}\label{subsec:dragonfly}

While the KCWI spectroscopy confirms that a patch of visible emission in the outer region of the HST image is, in fact, part of the galaxy, we turn to the Dragonfly Telephoto Array \citep{Abraham:2014} to investigate whether more (or even the full extent) of Ring 2 can be identified via deeper imaging. Dragonfly is optimized for low surface brightness observations, and is known for its discovery and characterization of ultra-diffuse galaxies and stellar halos in galactic outskirts. Using Dragonfly, we obtained $\sim400\times10$ minute exposures of \leda{} in both the sloan$-g$ and sloan$-r$ band filters. Details of the general observing and reduction strategy for Dragonfly can be found in \cite{Danieli:2020}. 

The $g$ band Dragonfly image is shown
in the top left panel of Figure \ref{fig:subtracted}.
As mentioned, \leda{} is in a region with fairly strong Galactic cirrus contamination (though the strongest bands appear to just miss the galaxy). We model the cirrus contamination using techniques discussed in Liu et al. (\textit{submitted}), where a more detailed technical description can be found; here we summarize the steps. 

First, the coadded images are sky subtracted following the procedures described in \cite{Liu:2023}. The methods therein are designed to mitigate the fact that traditional sky-background estimations (e.g., box averaging or spline fitting) would be biased by the presence of the cirrus itself. In short, it involves adopting a prior based on far infrared / sub-millimeter data from the Planck satellite, assuming the dust is optically thin on large scales and in thermal equilibrium, allowing the contributions of airglow and zodiacal light (among other less-contributing time-varying, large scale sky patterns) to be subtracted while preserving the details of the cirrus. 

Next, the coadded Dragonfly image is passed through the \texttt{mrf} code \citep{vanDokkum:2020}. This code performs \textit{multi-resolution filtering}, using higher resolution imaging (in this case, Legacy Surveys DR10 imaging from \cite{Dey:2019}) to construct a flux model of compact and high surface brightness sources in the images, which are then convolved with a kernel which matches them to Dragonfly imaging. This flux model is then subtracted, leaving behind (in principle) only low surface brightness, extended features, with ``low surface brightness'' being defined by the cutoff in magnitude used to construct the flux model in the higher resolution imaging. As a note, particular care is given to modeling the extended wings of the PSF using a non-static model given the high cirrus content; details of this procedure can be found in Liu et al. (\textit{submitted}). Finally, the cirrus structure is isolated by modeling the cirrus SED using Dragonfly imaging data and Planck dust models \citep{2014A&A...571A..11P} and applying a morphological filtering following the parameters of Liu et al. (\textit{submitted}). This final cirrus model, shown in the top right panel of Fig.\ \ref{fig:subtracted}, is then subtracted from the Dragonfly MRF image, with masked locations infilled using the \texttt{maskfill} code \citep{vanDokkum:2024}, which uses a robust median infilling procedure that iteratively fills masks using the information present at the mask edges.

In the bottom row of Figure \ref{fig:subtracted},  we show the Dragonfly $g$-band image at left (after MRF has been used to subtract the compact, high surface brightness sources) with masks overlaid where there were remaining residuals. At right we show the cirrus-subtracted final image (with those masks infilled using \texttt{maskfill}), with moderate smoothing with a $\sigma=1$ px applied. In both images, the white rectangular aperture shows the location of the KCWI pointing within which we detect $H\alpha$ at the redshift of \leda. As expected, this patch is particularly bright, which is why it was detectable in the Legacy and HST imaging. 

We see clear evidence for the presence of Ring 2 around the galaxy in both the original and residual image, strengthened by the structure's consistency with the expected axis ratio and position angle. In both bottom panels of Figure \ref{fig:subtracted}, we show a white ellipse with a manual ellipse match to the visible ring structure, with $b/a=0.75$ and $\theta=19$ degrees, consistent with the inner rings; a relatively large center offset from the galaxy center is required to reasonably-trace the structure. In addition to this direct modeling method, we also investigate whether the distribution of cirrus traced in the infrared (via WISE) shows any sign of the ring structure as an independent test; these results are presented in the Appendix, and in short we find no evidence for any similar looking structure in WISE, further motivating its extragalactic origin to the structure.

The signal-to-noise is not sufficient to attempt a ring-centroiding procedure as was carried out for the HST data, so we can only approximately estimate that Ring 2 has a semi-major axis of $\sim69.5''$, (assuming a $b/a=0.75$ and $\theta=19$); these parameters provide a reasonable match to the data (as shown in Figure \ref{fig:subtracted}), but have considerable uncertainty. Assuming these measurements for Ring 2, the discrepancy between the predicted and ``observed'' ring semi-major axis is $\sim$10\%. The relative uncertainty estimates for the centroided inner rings (while admittedly conservative) are on the order of $\sim4.5$\% on average, with a maximum-derived uncertainty of 7\%. We thus adopt a blanket uncertainty on this ring of 2 times that average (9\%); this is the errorbar reflected in Figure \ref{fig:ring-compare}. 

Given this uncertainty estimate for Ring 2, we conclude that, like the other identified rings, this ring likely does not present any strong tension with analytic theory. Indeed, it is the most remarkable of the agreements, being the oldest and most diffused, expanded ring detected in this work.

\begin{figure*}[htp!]
    \centering
    \includegraphics[width=\linewidth]{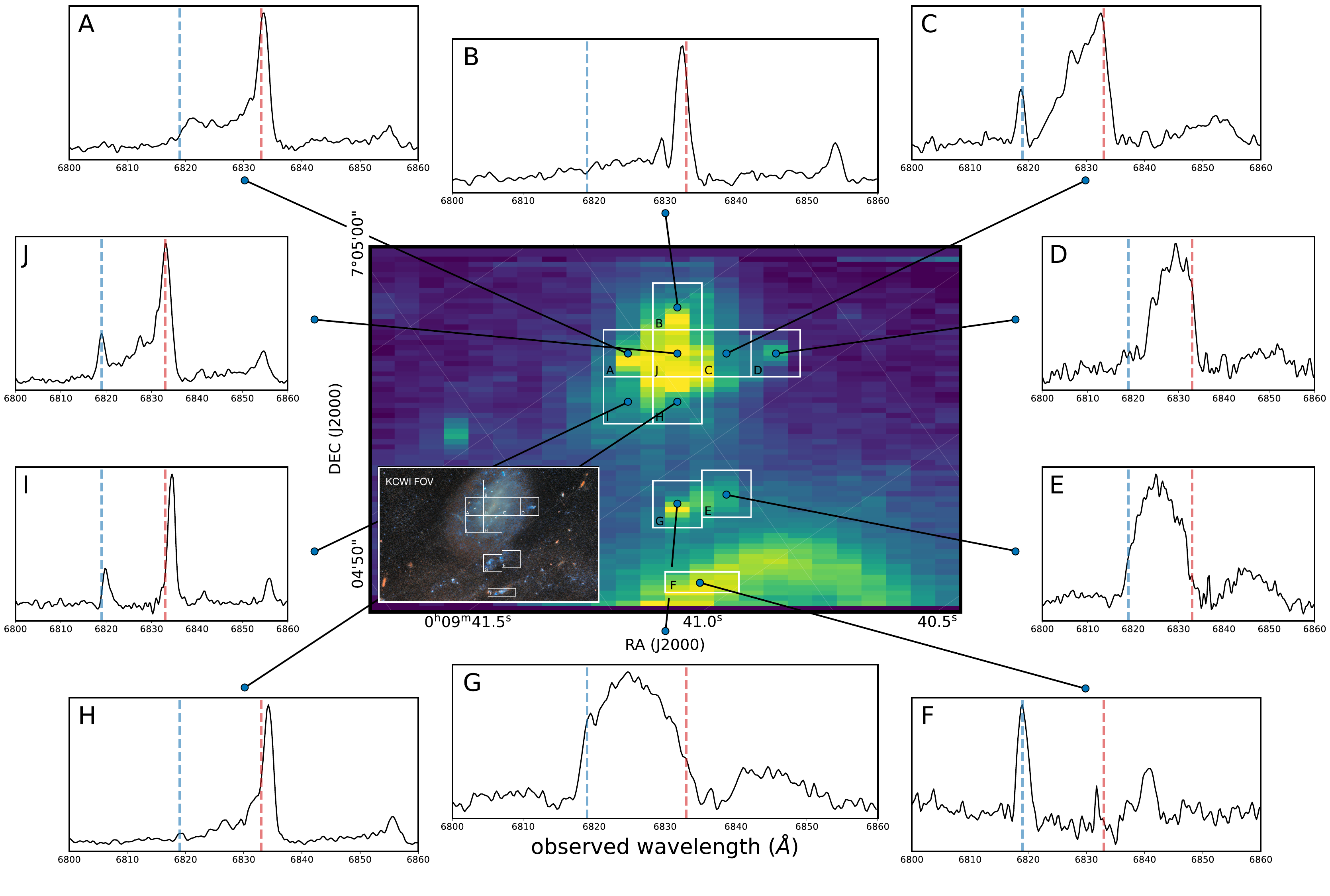}
    \caption{KCWI moment 0 map of the dwarf and ring of the Bullseye (center panel), with 1D spectra extracted from various apertures shown in panels A-J. To guide the eye, we show the regions targeted by each extraction over the ESA-created HST image in the corner of the center panel. In all spectra, a vertical blue dashed line marks the rough velocity of the Bullseye galaxy as traced by the narrow line emission in the ring feature in aperture F, while a dashed red line marks the rough velocity of the dwarf galaxy as traced by the narrow component of emission in the central pointing in aperture J. These extractions trace a disturbed and complex ionization and kinematic state for the gas in the dwarf and between the dwarf and the Bullseye. In particular, we note that broad ($\sigma>500$ km s$^{-1}$) components of emission exist both in relative spatial isolation (as in apertures D, E, F) or blended with narrow line emission (as in apertures A, H, J). The velocities of this gas span the full wavelength range between the dwarf and the Bullseye, and, critically, do not appear significantly \textit{outside} this velocity region. Paired with the non-Gaussian shape of the broad line emission, these spectra are highly suggestive of gas either being stripped from the dwarf and trailing behind, or gas liberated from \leda{} and moving toward the dwarf, or both. It is challenging to explain the observed lines (and line ratios) in any scenario in which the dwarf is not the impacting galaxy.}
    \label{fig:kcwi-dwarf}
\end{figure*}

\section{Identity of the Impactor}\label{sec:impactors}

Of key importance to follow-up modeling and simulations of the \leda{} system is the identification of the so-called ``impactor'' --- the galaxy which flew through the center of the Bullseye to trigger the rings. Because the impact must have occurred less than 1 Gyr ago, with impact velocities ranging from $\sim$500-2000 km s$^{-1}$, and because \leda{} is moderately face-on, the impactor must be within a reasonably small angular field around the Bullseye. 

Of the visible galaxies, three have (spatially) a reasonable likelihood of being the impactor; these are the three systems visible in Figure \ref{fig:hst-1}. We obtained Keck/LRIS longslit spectroscopy for both the blue, clumpy dwarf galaxy\footnote{Note that we will refer to this system as a dwarf colloquially hereafter due to its relative size and morphology, but its mass is not well known, and it may be more massive than it appears. Recall that the size of the Bullseye to scale is significantly larger than the Milky Way.} in the north-east of \leda{} as well as the small elliptical to the north-west. Their radial velocities with respect to the Bullseye are 525 and -385 km s$^{-1}$ respectively; therefore, both have velocities consistent with being part of the \leda{} system. 

Of the two, the blue, clumpy dwarf appears to be suggestive as a potential impactor; it is clearly disturbed, with numerous blue clumps of star formation (possibly triggered during the collision). Given the inclination of the system, this could place its orthogonal velocity with respect to \leda{} in the range simulations suggest is needed for CRG formation.

We investigate this possibility via a KCWI pointing centered on the dwarf galaxy and intervening space between it and the center of the Bullseye. Data were obtained using the Large slicer and BH3/RH2 gratings on October 2, 2024, in sub-arcsecond seeing. 

There are several observational signatures that could robustly identify the dwarf galaxy as the impacting system: 

\begin{itemize}
    \item Disturbed kinematics and atypical line ratios consistent with a recent merger-like event, and
    \item A trail of gas extending between the dwarf and the Bullseye; e.g., gas stripped from the dwarf or liberated from the host galaxy by the dwarf during its passage (or both).
\end{itemize}

In  Figure \ref{fig:kcwi-dwarf}, we present extracted spectra from the IFU observations, showing ten representative apertures across the dwarf, several clumps of intervening emission, and for reference, the ring of the Bullseye that is visible in-frame. We show the extraction regions over the moment 0 image computed in the region surrounding $H\alpha$, with an inset of the same pointing in the HST image to guide the eye. In each 1D spectrum, we indicate the 
velocity of the Bullseye, 11,731 km s$^{-1}$, in blue, and the velocity  of the dwarf, 12,257 km s$^{-1}$, in red. Note that this Bullseye velocity is $\sim100$ km s$^{-1}$ different than the value inferred from the ALFALFA HI data; we use the KCWI-measured velocity in this section as all KCWI pointings share the same internal calibration, meaning the relative velocity differences are robust.

It is clear from the remarkably disturbed and kinematically-complex structure of this system that it is not a typical dwarf galaxy. Of particular note are regions of extremely broad emission of $W \geq$500 km s$^{-1}$ in line width, corrected for redshift and instrumental broadening (apertures D, E, and G). Here we use $W$ as the full width of the feature; as they are non-gaussian, flattened, and have steep wings. These regions do not show narrow line emission and the emission spans velocities extending nearly-exactly from the dwarf velocity to the bullseye velocity, strongly suggesting its origin is related to the impact. Other apertures (e.g., A, C, H, J) show a mix of narrow line emission from the dwarf and a broad, blue-shifted ``shelf-like'' component. Note that on occasion due to projection effects, $H\alpha$ from the bullseye is also visible in these extractions, identifiable by the blue dashed vertical line. While not shown in the spectrum cutouts in Figure \ref{fig:kcwi-dwarf}, the [SII] doublet follows the same trend in broad and narrow components. 

Given these spectral properties, and in particular the fact that significant components of gas velocities ``fill'' the phase space between the Bullseye and the dwarf, we infer that the dwarf-like system seen near in projection to \leda{} is indeed the impactor which produced the ring system seen today. The obtained spectra would be challenging to explain in any other scenario, including AGN feedback or general starburst / outflows.

Based on this determination, we can estimate the time since the collision (and distance between the dwarf and \leda) under some specific assumptions about the interaction geometry. If we assume a perfectly head-on collision, then taking the measured 41.9 degree inclination of the system, we infer a total collision velocity of 705 km s$^{-1}$. Similarly, this assumption implies that the visible spatial separation between the dwarf and Bullseye center on-sky is entirely due to the inclination; thus, we can infer an intrinsic distance between \leda{} and the dwarf to be 40.6 kpc (the angular separation of 43.42$''$ corresponds to 36.42 kpc at the assumed distance of 173 Mpc). If we further assume that the dwarf has had a constant velocity since the collision, then we find that the collision should have occurred $\sim$56 Myr ago. 

This timescale is roughly consistent with predictions from simulations. While there have been no simulations to date which fully capture the characteristics of \leda{} in particular, many numerical simulations \citep[e.g.][]{Mapelli:2008,Fiacconi:2012,Smith:2012,Inoue:2021} find generically that the window of 50-150 Myr after the collision is where rings tend to be visible. The collision and ring-creation time can be independently constrained by the ring positions and expansion velocities. We have obtained followup KCWI imaging of all the rings and will investigate the dynamical state of the system and halo in more detail in a later work.

\section{A Pathway to GLSBs?}\label{sec:glsbs}

Beyond its immediate interest as a nine (inferred ten) ringed galaxy, the presence of faded, extended rings at large radii raise the interesting question of whether the end-state of \leda{} may in fact resemble known GLSBs such as Malin I. 

Such an end state has been predicted by simulations; \cite{Mapelli:2008} found that simulations attempting to mimic the Cartwheel galaxy (the most famous example of a CRG), after $\sim$1 Gyr, evolved into giant, low surface brightness disks with similar properties to Malin I. However, other studies \citep[e.g.,][]{Boissier:2016} have disfavored this channel, finding in some observations of Malin I and several other known GLSBs that a collision is an unlikely precursor to their current states.\footnote{Observational evidence used in \cite{Boissier:2016} disfavoring a ring origin include not seeing ``spokes'' as in the simulations and the Cartwheel galaxy; we do not see distinct spokes in this galaxy either.} To date, there has not been the inverse study; that is, observational evidence from post-collision CRGs which suggests they can (or cannot) actually evolve as \cite{Mapelli:2008} suggests. Further, there is still little consensus on the dominant pathways for GLSB creation, other than the agreed-upon maxim that entirely secular processes are generally unable in $\Lambda$CDM to create disks of the sizes and surface brightnesses seen. 

\leda{} may be providing the first direct observational evidence for the CRG-GSLSB evolutionary pathway. It is the first CRG for which it has been shown that extended, faded rings can lead to low-surface-brightness extended features at galactocentric radii similar to that of Malin I and other GLSBs. It is perhaps not unexpected that such an object has not yet been characterized, given the exceedingly short timescales over which an observer must ``catch'' a CRG in this multi-ring state, paired with the limited volume within which follow up imaging and spectroscopy at the required depths and resolution is possible. 

\begin{figure}
    \centering
    \includegraphics[width=\linewidth]{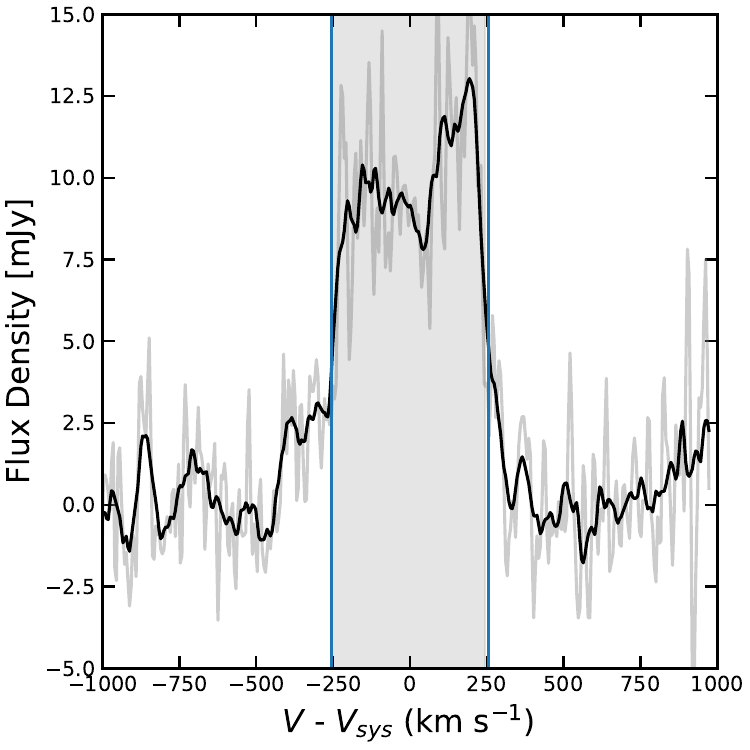}
    \caption{HI spectrum (gray) of \leda{} retrieved from the ALFALFA survey \citep{Haynes:2011}, with a boxcar smoothed version overlaid in black for clarity. The galaxy has a recognizable double-horned feature in HI with a typical rotation velocity on the order of 370 km s$^{-1}$ (correcting for inclination). Using the spectrum, we estimate the HI mass of \leda{} using the shaded region. }
    \label{fig:alfalfa}
\end{figure}

There is additional observational evidence that suggest \leda{} may be consistent with this picture: A hallmark of GLSBs is that they have elevated HI gas masses for their stellar masses compared to typical spirals. To characterize this parameter ($M_{\mathrm{HI}}/M_{*}$) for the Bullseye, we retrieve an integrated HI spectrum of \leda{} from the ALFALFA survey \citep{Haynes:2011,Haynes:2018}.\footnote{The spectrum was retrieved from the NASA/IPAC Extragalactic Database.} The spectrum, corrected for the galaxy's systemic velocity in HI, and for the measured redshift of $z=0.039$, is shown in Figure \ref{fig:alfalfa}, with the window within which we compute the mass marked.

We compute the total HI mass for \leda{} via 

\begin{equation}
    M_{\mathrm{HI}} = \frac{2.36 \times 10^{5}}{1+z} D_L^2 \int F dV, 
\end{equation}
where $D_L$ is in Mpc, velocities are in \kms{}, and $F_{HI}$ is in Jansky \kms{} \citep[e.g.,][]{Lutz:2017}, inferring a total HI mass of 
\begin{equation}
    M_{\mathrm{HI}} = (3.5 \pm 1.6) \times 10^{10}\; M_{\odot},
\end{equation}
or a $\log M_{\mathrm{HI}}$ of $10.54\pm 0.21$. Uncertainties in the HI mass were estimated in two ways; (1) via perturbation bootstrapping the measurement on the spectrum, and (2) via the technique of \cite{Doyle:2006}. The latter produced the (larger) uncertainties quoted; the bootstrapping estimated lower uncertainties of $\sim0.13\times10^{10}$ $M_{\odot}$. As a note; while the beam size for ALFALFA is large and for \leda{} contains the three other galaxies visible in Figure \ref{fig:hst-1}, we assume they do not contribute appreciably to the computed mass.

Next, we estimate the stellar mass of the Bullseye using $g$ and $r$ band imaging from the Legacy survey DR9. Reduced, background-subtracted images were retrieved from the Legacy service on 13 February, 2024. We measure the flux of \leda{} in nanomaggies from the two images using a photometric aperture which extends to the outer star-forming ring. Fluxes were converted to AB magnitudes using the common imaging zeropoint of 22.5, that is, $m = 22.5 -2.5\log(F)$. The legacy magnitudes were shifted to approximate SDSS via $g_{\mathrm{SDSS}} = g_{l}+0.09$ and $r_{\mathrm{SDSS}} = r_l + 0.1$, following \cite{Mao:2021}, from which the shift was determined using the region of overlap between the surveys. Extinction corrections for each band were retrieved from the NASA Extragalactic Database.

We compute the absolute magnitude in the $r$-band using a luminosity distance of 173.1 Mpc for redshift 0.039, and apply a $k$-correction of 0.04 computed following \citep{Chilingarian:2010}. The stellar mass was then estimated following \cite{Mao:2021},
\begin{equation}
    \log(M_*/M_{\odot}) = 1.254 + 1.098(g-r) - 0.4m_r
\end{equation}
where $m_r$ is the absolute magnitude in the $r$ band. This equation applies a moderate (0.3 dex) correction to the mass-to-light vs. color relation presented in \cite{Bell:2003} to convert from a diet Salpeter to Kroupa IMF (0.15 dex) and to better match recent samples (0.15 dex). The conversion assumes a solar $r$-band absolute magnitude of 4.65. 

Using this method, we find a stellar mass of $\log(M/M_{\odot}) = 10.77$ for \leda{}. We assume a blanket systematic error of 0.2 dex. As a note, this calculation neglects any stellar mass that may be present in the outermost rings, but such a component is likely negligible. We similarly apply no aperture correction, assuming the aperture enclosing the outer SF ring includes all the relevant stellar mass in the system. 

\begin{figure*}
    \centering
    \includegraphics[width=0.8\linewidth]{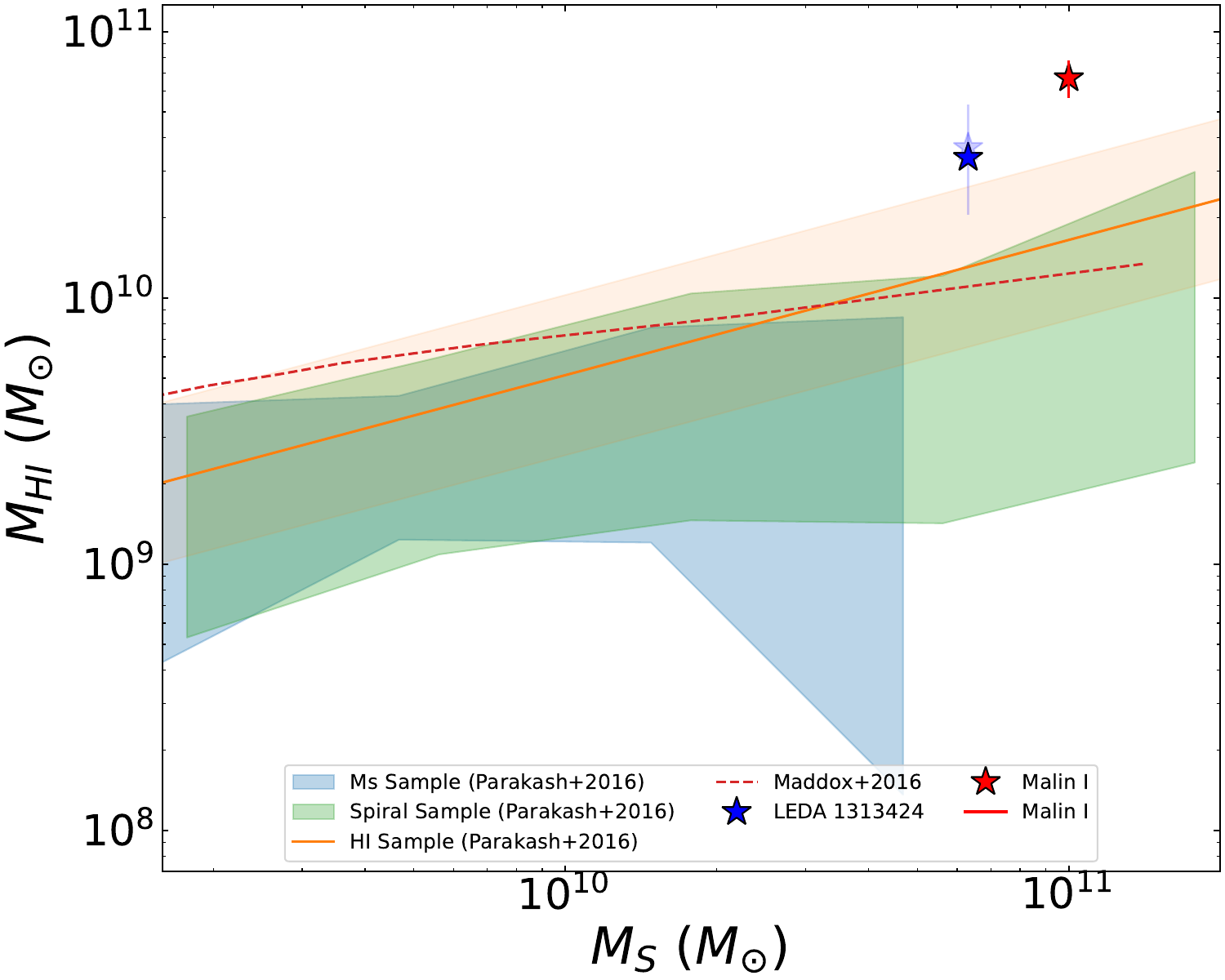}
    \caption{Measured HI/stellar mass relations used in \cite{Parkash:2018}, from samples selected using stellar mass (blue shaded region), HI properties (orange line and shaded region), and morphology (spirals; green shaded region), as well as a relation from \cite{Maddox:2016} (red dotted line). \leda{} and Malin I are shown in this space as individual stars. We find that both galaxies are relatively rich in HI for their stellar masses, $\sim1\sigma$ or more above the known relations.  }
    \label{fig:hi-compare}
\end{figure*}

In Figure \ref{fig:hi-compare}, we show the measured HI mass vs stellar mass for \leda{} alongside Malin I \citep[from][]{Lelli:2010,Boissier:2016}, compared to several HI-$M_{s}$ relations derived in \cite{Parkash:2018}. Including the blueward tail in the mass estimate produces the lower-opacity star. \leda{} is elevated from samples of typical spiral galaxies for its stellar mass (sitting $>1\sigma$ above the relation), though not to as strong a degree as Malin I. 

While significant further study is required to comprehensively assess the viability of the CRG-to-GLSB pathway, it is certainly striking that these otherwise independent lines of evidence place \leda{} in contention to be a future GLSB. Detailed simulations matching the scenario which created this system can help explore this possibility. While this system alone cannot constrain the \textit{rates} at which CRGs may follow this kind of pathway, which likely depends heavily on the parameters of the progenitor galaxy, the impacting galaxy, and the collision dynamics, the Bullseye does for the first time observationally support the possibility that this pathway is viable. 

\section{Summary and Conclusions}
We present multiwavelength observations of \leda, the first Collisional Ring Galaxy captured in the narrow moment of the post collision timeframe during which nine rings are visible (with ten rings being inferred). Combining HST and Dragonfly imaging with Keck/KCWI spectroscopy, we characterize the ring structures in the galaxy and discuss implications for its future evolution.
\begin{itemize}
    \item We measure the radii of all detected ring structures and find remarkable agreement with the (relatively simple) analytic ring theory presented in \citet{Struck:2010}. 
     \item We identify patches of emission in the HST imaging at very large galactocentric radii ($r\sim70$ kpc) that appear to belong to the faded second ring; we confirm the association of this material with the galaxy via KCWI spectroscopy (which detects numerous clumps of $H\alpha$ at the appropriate redshift) as well as via Dragonfly low surface brightness imaging, which appears to capture the full extent of the ring. 
    \item While we do not attempt to robustly fit this outer ring, we find that estimates for its properties place it $\sim$10\% larger than expected by the ring theory, in reasonable agreement.
    \item We investigate the clumpy blue system near to the Bullseye in projection with Keck/KCWI and find strong evidence that it is the impactor, in particular, the presence of significant gas spanning velocities between the two systems.
    \item We discuss the hypothesis that CRGs can evolve into GLSBs as their rings expand and fade, a mechanism that has been put forward before but has not been observationally verified; \leda{} may serve as a transition object in which both rings \textit{and} material at large radii demonstrate the possibility of this pathway.
    \item Finally, we estimate the stellar mass (from photometry) and neutral hydrogen (HI) mass (from ALFALFA spectroscopy) and show that in line with GLSB properties, the Bullseye has highly elevated HI mass for its stellar mass, providing another line of evidence that this system is approaching consistency with GLSBs. 
\end{itemize}

We additionally present surface brightness profiles (and the associated size determinations) for \leda{} in the Appendix. 

\vspace{20pt}

IP would like to thank Chenghan Hsieh for helpful discussions. 

The Legacy Surveys consist of three individual and complementary projects: the Dark Energy Camera Legacy Survey (DECaLS; Proposal ID \#2014B-0404; PIs: David Schlegel and Arjun Dey), the Beijing-Arizona Sky Survey (BASS; NOAO Prop. ID \#2015A-0801; PIs: Zhou Xu and Xiaohui Fan), and the Mayall z-band Legacy Survey (MzLS; Prop. ID \#2016A-0453; PI: Arjun Dey). 

DECaLS, BASS and MzLS together include data obtained, respectively, at the Blanco telescope, Cerro Tololo Inter-American Observatory, NSF’s NOIRLab; the Bok telescope, Steward Observatory, University of Arizona; and the Mayall telescope, Kitt Peak National Observatory, NOIRLab. Pipeline processing and analyses of the data were supported by NOIRLab and the Lawrence Berkeley National Laboratory (LBNL). 

The Legacy Surveys project is honored to be permitted to conduct astronomical research on Iolkam Du’ag (Kitt Peak), a mountain with particular significance to the Tohono O’odham Nation.

NOIRLab is operated by the Association of Universities for Research in Astronomy (AURA) under a cooperative agreement with the National Science Foundation. LBNL is managed by the Regents of the University of California under contract to the U.S. Department of Energy.

This project used data obtained with the Dark Energy Camera (DECam), which was constructed by the Dark Energy Survey (DES) collaboration. 

Funding for the DES Projects has been provided by the U.S. Department of Energy, the U.S. National Science Foundation, the Ministry of Science and Education of Spain, the Science and Technology Facilities Council of the United Kingdom, the Higher Education Funding Council for England, the National Center for Supercomputing Applications at the University of Illinois at Urbana-Champaign, the Kavli Institute of Cosmological Physics at the University of Chicago, Center for Cosmology and Astro-Particle Physics at the Ohio State University, the Mitchell Institute for Fundamental Physics and Astronomy at Texas A\&M University, Financiadora de Estudos e Projetos, Fundacao Carlos Chagas Filho de Amparo, Financiadora de Estudos e Projetos, Fundacao Carlos Chagas Filho de Amparo a Pesquisa do Estado do Rio de Janeiro, Conselho Nacional de Desenvolvimento Cientifico e Tecnologico and the Ministerio da Ciencia, Tecnologia e Inovacao, the Deutsche Forschungsgemeinschaft and the Collaborating Institutions in the Dark Energy Survey. 

The Collaborating Institutions are Argonne National Laboratory, the University of California at Santa Cruz, the University of Cambridge, Centro de Investigaciones Energeticas, Medioambientales y Tecnologicas-Madrid, the University of Chicago, University College London, the DES-Brazil Consortium, the University of Edinburgh, the Eidgenossische Technische Hochschule (ETH) Zurich, Fermi National Accelerator Laboratory, the University of Illinois at Urbana-Champaign, the Institut de Ciencies de l’Espai (IEEC/CSIC), the Institut de Fisica d’Altes Energies, Lawrence Berkeley National Laboratory, the Ludwig Maximilians Universitat Munchen and the associated Excellence Cluster Universe, the University of Michigan, NSF’s NOIRLab, the University of Nottingham, the Ohio State University, the University of Pennsylvania, the University of Portsmouth, SLAC National Accelerator Laboratory, Stanford University, the University of Sussex, and Texas A\&M University.

BASS is a key project of the Telescope Access Program (TAP), which has been funded by the National Astronomical Observatories of China, the Chinese Academy of Sciences (the Strategic Priority Research Program “The Emergence of Cosmological Structures” Grant \# XDB09000000), and the Special Fund for Astronomy from the Ministry of Finance. The BASS is also supported by the External Cooperation Program of Chinese Academy of Sciences (Grant \# 114A11KYSB20160057), and Chinese National Natural Science Foundation (Grant \# 12120101003, \# 11433005).

The Legacy Survey team makes use of data products from the Near-Earth Object Wide-field Infrared Survey Explorer (NEOWISE), which is a project of the Jet Propulsion Laboratory/California Institute of Technology. NEOWISE is funded by the National Aeronautics and Space Administration.

The Legacy Surveys imaging of the DESI footprint is supported by the Director, Office of Science, Office of High Energy Physics of the U.S. Department of Energy under Contract No. DE-AC02-05CH1123, by the National Energy Research Scientific Computing Center, a DOE Office of Science User Facility under the same contract; and by the U.S. National Science Foundation, Division of Astronomical Sciences under Contract No. AST-0950945 to NOAO.

\appendix 

\begin{figure*}
    \centering
    \includegraphics[width=\linewidth]{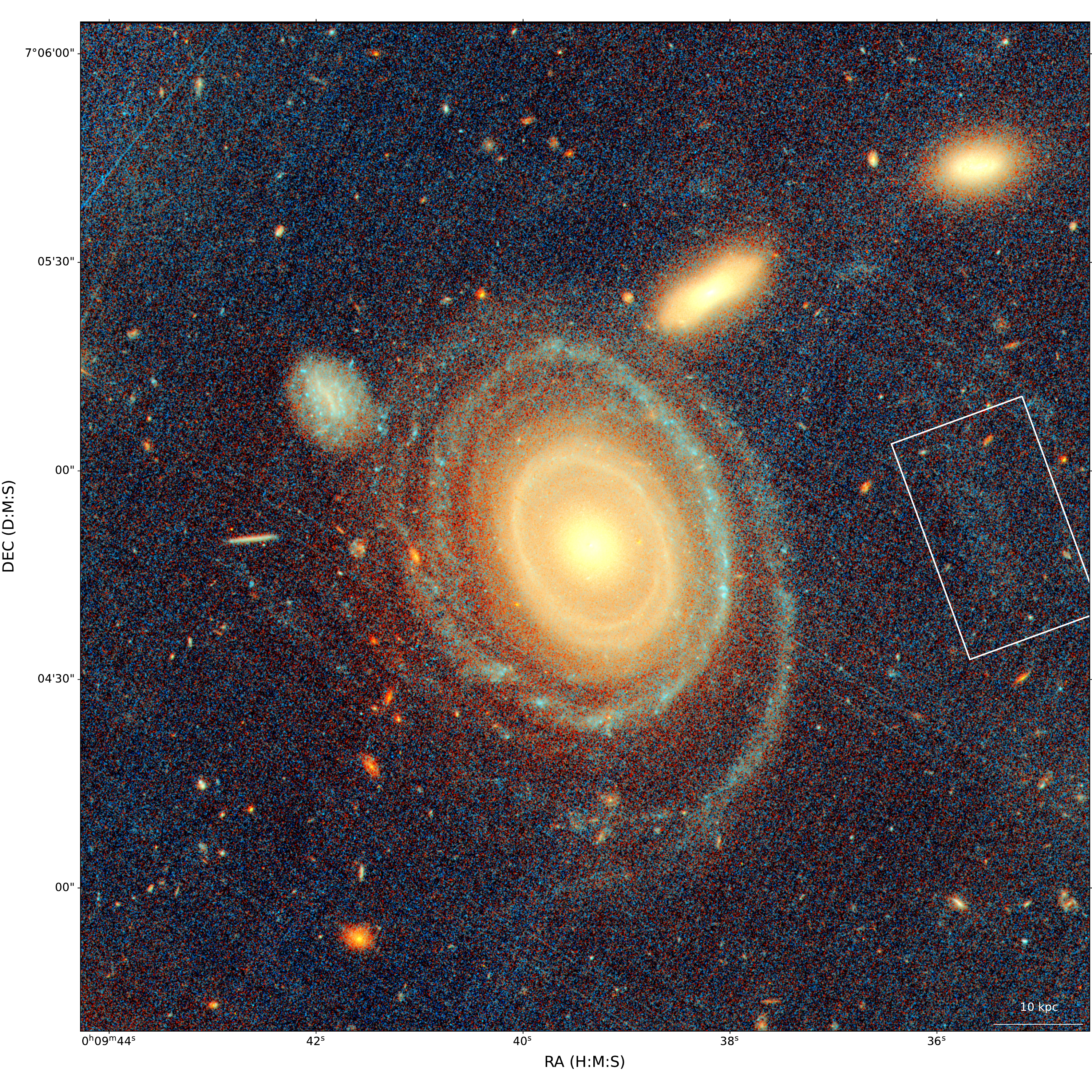}
    \caption{Composite HST color image of \leda{} (the ``Bullseye'') created by the authors using the F475W and F814 imaging, and placed onto the world coordinate system. Here we demarcate a white box over the patch of blue emission seen to the west of the galaxy, to indicate the pointing for KCWI, which we use to spectroscopically confirm the redshift of this emission. All measurements presented in this work were made on the science-calibrated images used to produce this composite. }
    \label{fig:hst-1}
\end{figure*}

In Figure \ref{fig:hst-1}, we show the authors' combination of the two-band HST imaging; all measurements made herein were derived directly from the science data and not the imaging in Figure \ref{fig:esa}. 

\section{Size Measurements and Surface Brightness Profiles}

To aid in the overall characterization of \leda{} as well as to provide context for the discussion in Section \ref{sec:glsbs}, we use ancillary data here to determine several basic properties of the galaxy, including estimates for its stellar mass and neutral hydrogen content.

The size of \leda{} is particularly important when considering the dynamical state of the rings and subsequent evolution. We compute both traditional size measured (e.g., $r_e$, $r_{80}$) as well as note the size of various noted features of the galaxy (e.g., exterior rings). 

\begin{figure*}
    \centering
    \includegraphics[width=\linewidth]{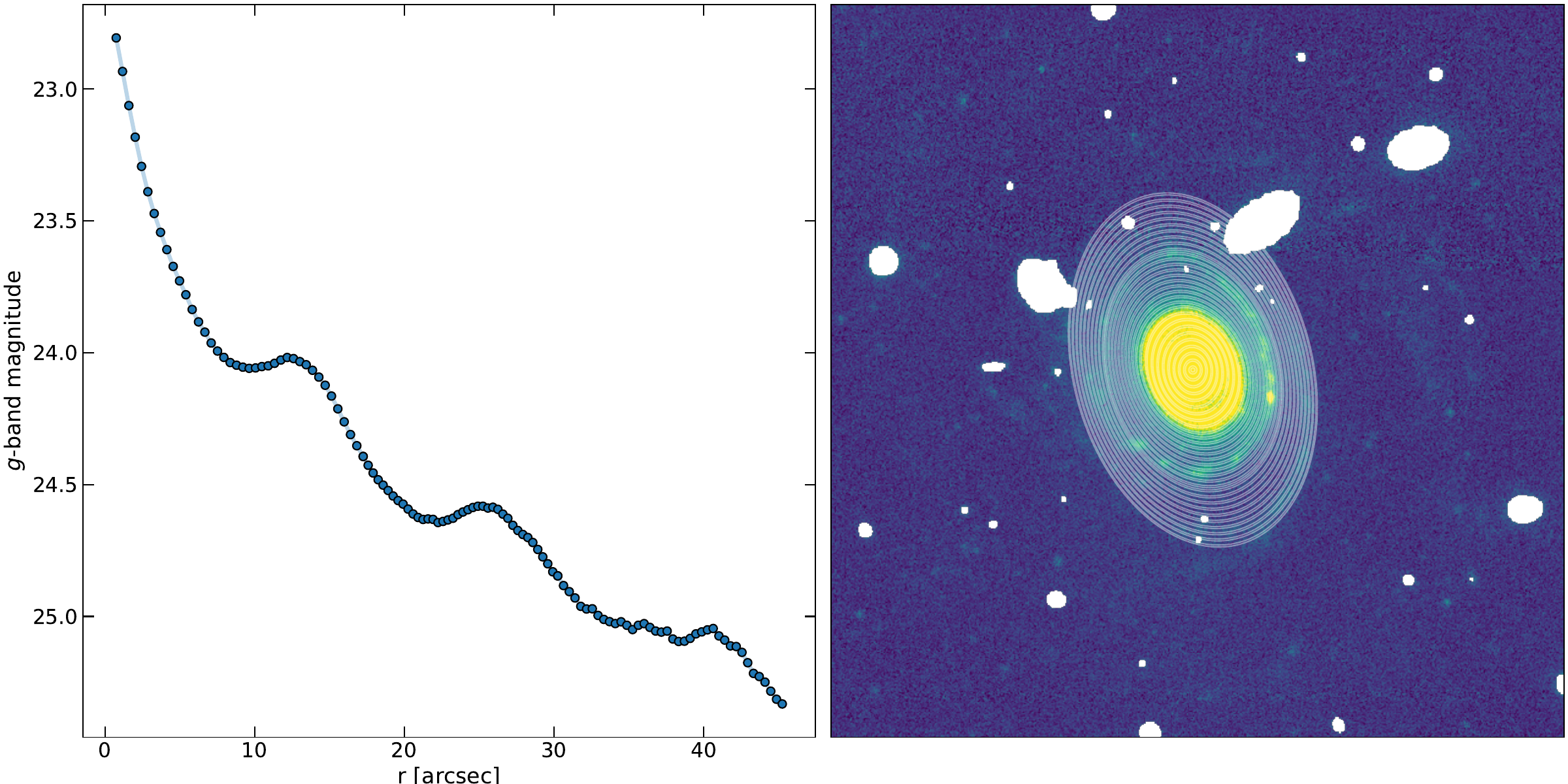}
    \caption{Surface brightness profile in the $g$-band for \leda{}. We find that the strongest rings (the red ring set and two blue rings) are easily discernible in the profile. Total flux is measured by integrating over the measured profile (and compared with the measured photometry used to derive the stellar mass). We find $r_e=19.75$ kpc and $r_{80}=36.14$ kpc.}
    \label{fig:sb-profile}
\end{figure*}

To compute traditional sizes, we construct a surface brightness profile using elliptical annuli extending from the core of the galaxy out to ring 3. Each annulus has a width of $\sim$0.38 arcseconds. Here we allow the precise position angle to vary; in practice, we found $20 < PA < 25$ degrees across the full galaxy. The nearby dwarf and elliptical systems were masked. The computed surface brightness profile is shown in Figure \ref{fig:sb-profile}. 

Based on the profiles, we find a half-light radius of 19.75 kpc, with an $r_{80}$ of 36.14 kpc. This places the galaxy on the upper end of the size-mass relation at redshift $z\sim0$ \citep[e.g.,][]{Lange:2015}, with $\sim2x$ larger half-light radius than similarly massed disks (or ellipticals). 

Additionally, the distance along the major axis to which we detect emission in the Dragonfly imaging is $100''$, corresponding to a total extent of $\sim77$ kpc. 

\section{Cirrus Comparison with WISE W3}

The optically-emitting galactic cirrus that was modeled and subtracted from the Dragonfly imaging also emits in the infrared, and is detected in WISE imaging bands (W1-W4). Here we briefly investigate the cirrus distribution as traced by dust infrared emission, to search for any sign of the structure we identify as Ring 2. Presence of the structure in WISE would complicate the ring interpretation, as if the structure is indeed extragalactic in origin and is a faded, recently-starforming ring, there is no expectation to see it in WISE, which we retrieve from the NASA/IPAC Infrared Science Archive (IRSA). Here we select W3 our comparison band, as W4 has an appreciably larger beam size, while in the W1 and W2 bands much of the cirrus is already removed by sky subtraction. We note that the PSF of the WISE imaging (FWHM$\sim$7.4$''$ in W3)  is sufficient to detect the ring-structure seen, if it exists.

\begin{figure*}[!bht]
    \centering
    \includegraphics[width=\linewidth]{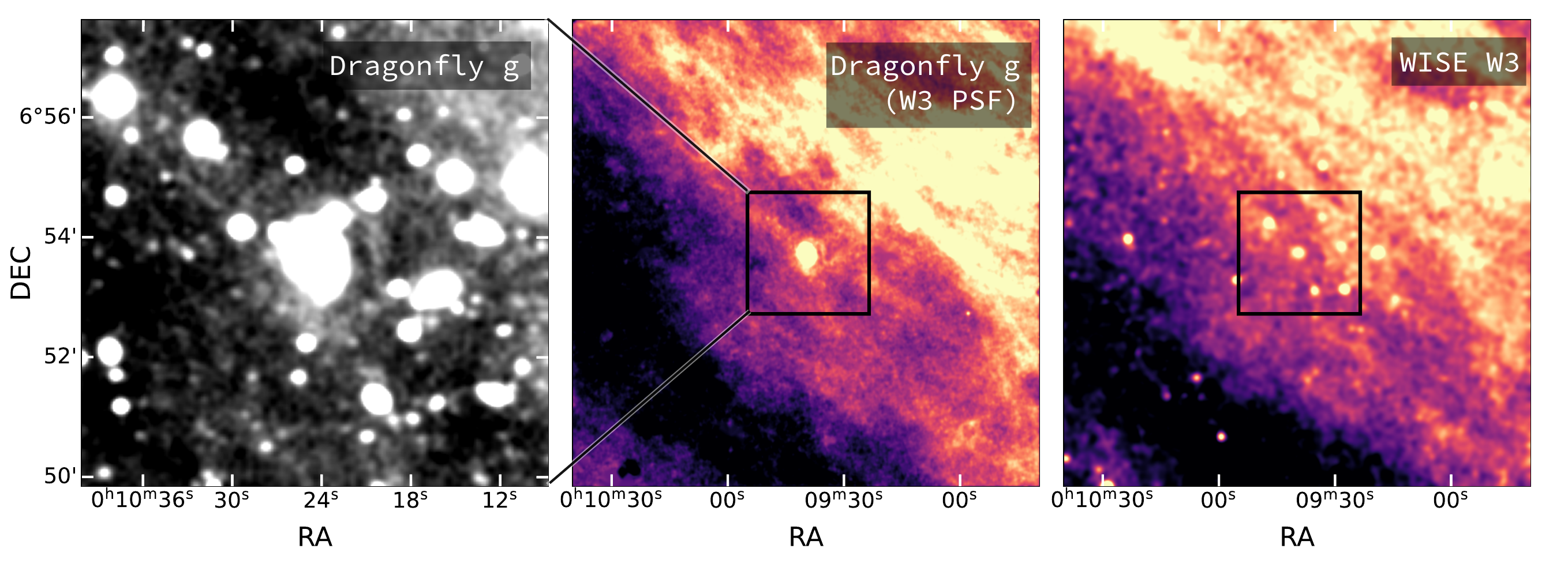}
    \caption{Smoothed Dragonfly $g$-band cutout of \leda{} (left), along with the same image in a larger field of view PSF-matched to WISE W3 (center) and and (smoothed) WISE (W3) band imaging (right) of the same region. The box in the center and right panel shows the field of view of the left hand cutout. The WISE data provides an independent probe of the galactic cirrus distribution, and finds no evidence for the ring-like structure seen in Dragonfly (though we note there is subtlety in comparing cirrus probes at different wavelengths, as they trace different mixtures of the dust grain distribution). If the ring structure is indeed extragalactic, it would not be expected to appear in the infrared map, thus providing additional evidence that the cirrus modeling and subtraction presented in Figure \ref{fig:subtracted} is legitimately identifying a ring.}
    \label{fig:wise}
\end{figure*}

 In Figure \ref{fig:wise}, we show a gaussian-smoothed version of the reduced Dragonfly $g$-band imaging (left) along with a wider view of the region that has been PSF-matched to WISE W3 (center) and the WISE imaging itself (right, smoothed to guide the eye). 

The WISE data clearly tracks a similar overall dust distribution, though there are subtle differences likely due to their tracing of different dust grain size distributions and populations. That said, we do not see any overdensity or structure around the galaxy location in any of the WISE bands, nor any indication of an equivalent structure as seen in Dragonfly. Based on this, we conclude that the structure seen in Dragonfly is indeed extragalactic, and that the cirrus modeling is adequately capturing the distribution of cirrus in the subtraction.

\bibliography{references}{}
\bibliographystyle{aasjournal}

\end{document}